\documentclass{article}
\usepackage{spconf,amsmath,graphicx}
\usepackage{amsfonts}
\usepackage{booktabs}
\usepackage{tipa}
\usepackage{setspace}

\usepackage{cite}
\usepackage{url}

\newcommand{\wt}[1]{\textipa{/#1/}}

\title{CREATING PERSONALIZED SYNTHETIC VOICES FROM ARTICULATION IMPAIRED SPEECH USING AUGMENTED RECONSTRUCTION LOSS}
\name{Yusheng Tian,  Jingyu Li, Tan Lee}
\address{Department of Electronic Engineering, The Chinese University of Hong Kong, Hong Kong SAR}
\begin{document}

\begin{figure*}
  \centering
  \includegraphics[width=1.0\linewidth]{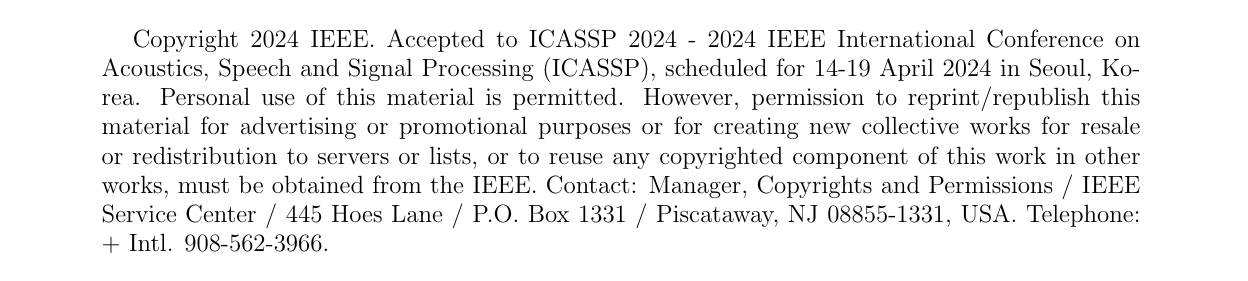}
\end{figure*}


\abovedisplayshortskip=0pt
\belowdisplayshortskip=0pt
\abovedisplayskip=4pt
\belowdisplayskip=4pt
\maketitle
%
\begin{abstract}
\begin{spacing}{0.95}
This research is about the creation of personalized synthetic voices for head and neck cancer survivors. It is focused particularly on tongue cancer patients whose speech might exhibit severe articulation impairment. 
Our goal is to restore normal articulation in the synthesized speech, while maximally preserving the target speaker’s individuality in terms of both the voice timbre and speaking style. This is formulated as a task of learning from noisy labels. We propose to augment the commonly used speech reconstruction loss with two additional terms.
The first term constitutes a regularization loss that mitigates the impact of distorted articulation in the training speech. The second term is a consistency loss that encourages correct articulation in the generated speech. 
These additional loss terms are obtained from frame-level articulation scores of original and generated speech, which are derived using a separately trained phone classifier. Experimental results on a real case of tongue cancer patient confirm that 
the synthetic voice achieves comparable articulation quality to unimpaired natural speech, while effectively maintaining the target speaker's individuality. Audio samples are available at https://myspeechproject.github.io/ArticulationRepair/.
\end{spacing}

\end{abstract}

\section{Introduction}
\label{sec:intro}
\begin{spacing}{0.95}

Tongue cancer is a prevalent form of head and neck cancer. Its incidence rate has been rising in recent years \cite{rising}. For advanced or recurrent tongue cancer, surgical intervention may involve the removal of both the tongue and larynx \cite{quality1,quality2}. As a consequence, the patient would lose voice and speaking ability permanently. Voice is not only an important means of communication, but also an integral part of a person’s identity. 
Personalized text-to-speech (TTS) was proposed to enable people with vocal disabilities to communicate using their own voices \cite{voicebank}. It was shown that using personalized TTS as an alternative communication method can significantly improve the quality of life of laryngectomees \cite{voca_improve_qol}.

Personalized TTS models are trained with natural speech from the target speaker. A major challenge in creating personalized synthetic voices for tongue cancer survivors is that the training speech is often impaired. Both the tumor and the treatment process could cause damages to the tongue, resulting in impaired speech production. Such impairments are 
most commonly at the articulation level \cite{impairment, symptom1}. TTS models trained with articulation-impaired speech would generate synthetic speech that expectantly contains similar types of impairment and may not meet the intelligibility requirement for communication. This motivates the present study on creating personalized synthetic speech from articulation-impaired training data. Our goal is to restore normal articulation in the synthesized speech while maximally maintaining the target speaker’s individuality. We consider the target speaker’s individuality to be well maintained if both the voice timbre and good aspects of the original speaking style are kept.


From the machine learning perspective, the above goal can be formulated as a task of learning from noisy labels\cite{noisylabel1, noisylabel2, noisylabel3, noisylabel4}. This is justified by the fact that the degree of articulation impairment in the speech of a tongue cancer patient varies across different types of speech sounds \cite{halpern2020detecting}: some sounds remain largely unaffected and can be viewed as having clean labels, while those with distorted articulation are with noisy labels. Inspired by the re-weighting approach \cite{reweighting1, reweighting2} and the consistency constraint approach \cite{reconstruction} developed in studies of learning with noisy labels, we propose to augment the conventional speech reconstruction loss in TTS model training with two additional terms. The first term is a regularization loss that mitigates the negative impact of distorted articulation in training speech. The second term is a consistency loss that promotes accurate articulation in the output speech. Specifically, a separately trained phone classifier is incorporated during training to provide frame-level articulation scores for both original and generated speech. The articulation score of original speech is used as the re-weighting criteria to derive the regularization loss. The articulation score of generated speech quantifies the inconsistency between the phone classifier and the TTS model, representing the consistency loss.

The proposed approach is validated on a real patient case, the same as the one reported in \cite{tian23b_interspeech}. A personalized synthetic voice is built for a female Cantonese speaker, who was advised to undertake laryngectomy for recurrent tongue cancer. 
The patient already underwent partial-glossectomy six years ago, and about 3/4 of her tongue was removed by surgical operation. As a result, she had difficulties in producing certain speech sounds. 
The synthetic voice is developed from the articulation-impaired speech of this patient using the augmented reconstruction loss. Objective and subjective evaluations are carried out to demonstrate the effectiveness of the augmented reconstruction loss in restoring normal articulation and maintaining the target speaker's individuality in the generated speech.
\end{spacing}
\vspace{-0.7em}

\section{Related work}
\begin{spacing}{0.95}

\label{sec:related_work}
\vspace{-0.3em}

\subsection{Voice reconstruction from impaired speech}
The problem of voice reconstruction from impaired speech was tackled in a few previous studies focusing on dysarthric speech. In the Voicebank project \cite{voicebank, creer2013building, veaux2012using}, it was proposed to repair speech by substituting impaired speech features with that of an average healthy voice for training statistical parametric model. More recently, a two-step approach was adopted to develop neural speech synthesizers for people with dysarthria \cite{vc1, vc2}. A TTS model is first trained with unimpaired speech of a healthy speaker. Subsequently, voice conversion is applied to transform the voice characteristics of the synthesized speech toward the target impaired speaker. Such VC-based approach may be sub-optimal for the specific case being considered in this study. With voice conversion, 
only the voice timbre is maintained and other speech characteristics of the target speaker are ignored, leading to a loss of speaker individuality.

In \cite{tian23b_interspeech}, we developed an approach to alleviate impaired speech articulation in synthesized speech with guided diffusion models. An external phone classifier was incorporated at the inference stage to identify unaffected time frames in the generated speech and guide the generation process to repair those affected ones. In this way, good aspects of the target speaker's speech characteristics are preserved. However, experimental results revealed a gap in the effectiveness of repairing distorted articulation compared to the VC approach. We speculate this to be related to inaccurate assessment by the phone classifier, which has to evaluate noisy speech during the reverse diffusion process. In contrast, in this work the phone classifier takes clean speech as input, therefore more accurate assessment from the phone classifier is expected.
\vspace{-0.5em}


\subsection{Learning from noisy labels}

The problem of learning from noisy labels has been extensively studied \cite{noisylabel1, noisylabel2, noisylabel3, noisylabel4}. The proposed regularization loss is related to the importance re-weighting approach in this field \cite{reweighting1, reweighting2}, where samples with corrupted labels are assigned small weights based on certain criteria. We utilize the articulation scores by the external phone classifier as the re-weighting criteria. The proposed consistency loss is inspired by the work of \cite{reconstruction}, which incorporates the notion of consistency into the training objective to handle noisy labels. In our work, consistency refers to the agreement between the phone classifier and the TTS model regarding the content of the generated speech. A similar loss has also been employed in \cite{crosslingual} to reduce foreign accent in cross-lingual voice conversion .
\end{spacing}
\vspace{-0.2em}

\section{Augmented reconstruction loss}
\vspace{-0.2em}

\begin{spacing}{0.98}
\begin{figure}[t]
  \centering
  \includegraphics[width=1.0\linewidth]{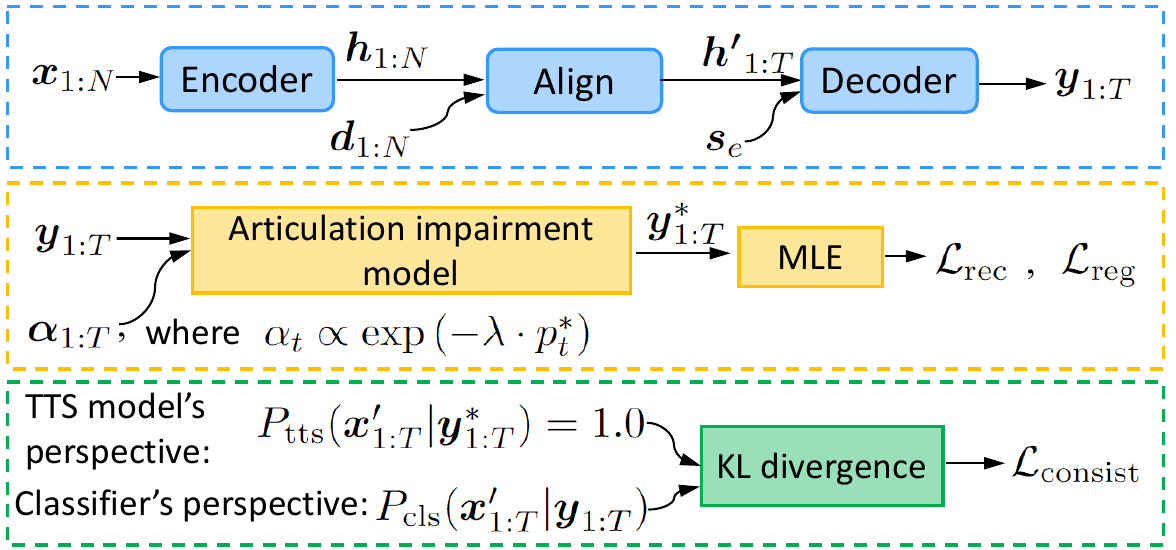}
  \caption{Overall framework of the TTS acoustic model.}
  \label{fig:framework}
  \vspace{-1.0em}
\end{figure}

A neural TTS system typically consists of two components: (1) an acoustic model that converts the input tokens into compact acoustic features; and (2) a vocoder that translates the acoustic features into a time-domain speech waveform. The present study is focused on the acoustic model as illustrated by the top block in Figure \ref{fig:framework}. The input tokens include phonemes in the language concerned and a silence token. The acoustic model is trained to generate a Mel-spectrogram \(\boldsymbol{y}_{1:T}\) that is close to the ground-truth \(\boldsymbol{y}^*_{1:T}\), given an input token sequence \(\boldsymbol{x}_{1:N}\). The input tokens are first encoded into hidden states \(\boldsymbol{h}_{1:N}\) through the encoder, and then expanded in time to align with \(\boldsymbol{y}^*_{1:T}\) by replicating each element according to its duration. The phoneme duration \(\boldsymbol{d}_{1:N}\) is obtained by forced alignment at the training phase and predicted by a separate duration model at the inference stage. The time-expanded hidden representations \(\boldsymbol{h}'_{1:T}\) are concatenated with the speaker embedding \(\boldsymbol{s}_e\) and passed to the decoder for Mel-spectrogram generation. 

The acoustic model is typically trained by minimizing the \(L_1\) and/or \(L_2\) loss between the ground-truth Mel-spectrogram \(\boldsymbol{y}^*\) and the predicted one \(\boldsymbol{y}\) \cite{tacotron2, durian}. When the training speech contains articulation impairment, synthetic speech generated by the trained model would contain similar types of impairment. To alleviate this problem, an augmented reconstruction loss is introduced. It incorporates two additional loss terms, namely regularization loss and consistency loss.
\vspace{-0.5em}


\subsection{Regularization loss}
As shown in the middle block in Figure \ref{fig:framework}, the regularization loss is derived by explicitly modelling articulation impairment and then applying maximum likelihood estimation.

Let \(\boldsymbol{y}_{1:T}\) be the Mel-spectrogram with unimpaired articulation generated by the acoustic model. We introduce an articulation impairment model, which transforms the Mel-spectrogram of normal speech into that of impaired speech:
%
\small
\begin{align}
P(\boldsymbol{y}^*_{1:T} | \boldsymbol{y}_{1:T})\propto \prod_{t=1}^T \Vert\boldsymbol{y}^*_t-\boldsymbol{y}_t\Vert^{\alpha_t}_2 \cdot\exp\left(-\frac{\Vert\boldsymbol{y}^*_t-\boldsymbol{y}_t\Vert_2^2}{2\sigma^2}\right)
\label{equation:impair},
\end{align}
%
\normalsize
where \(\alpha_t>0\) controls the severity of articulation impairment at the \(t^{th}\) speech frame. The variance \(\sigma^2\) is a constant term over time. A larger value of \(\alpha_t\) indicates more severe impairment, as it makes the mode of the distribution deviate further from unimpaired speech \(\boldsymbol{y}_t\).

The severity of impairment can be inferred using a separately trained phone classifier. We choose to set the values of \(\boldsymbol{\alpha}_{1:T}\) as follows:
\begin{align}
\alpha_t \propto \exp\left(-\lambda \cdot p^*_t\right) \text{ for } t=1 \cdots T
\label{equation:alpha},
\end{align}
where \(p^*_t\) denotes the posterior for the \(t^{\text{th}}\) phoneme label given the  observation \(\boldsymbol{y}^*_{1:T}\). The parameter \(\lambda>0\) controls the rate at which \(\alpha_t\) decreases as the posterior of the ground-truth phoneme increases.

Applying maximum likelihood estimation using the results from equations (\ref{equation:impair}) and (\ref{equation:alpha}), the following loss function can be obtained, where \(\beta>0\) is a hyper-parameter. 
\small
\begin{align}
\underbrace{\sum_{t=1}^T\Vert \boldsymbol{y}^*_{t} - \boldsymbol{y}_{t}\Vert_2^2}_{\mathcal{L}_{\text{rec}}} + \beta\cdot\underbrace{\sum_{t=1}^T -\exp\left(-\lambda \cdot p^*_t\right)\log\Vert\boldsymbol{y}^*_{t} - \boldsymbol{y}_{t}\Vert}_{\mathcal{L}_{\text{reg}}}
\label{equation:mle}
\end{align}
\normalsize
It consists of a conventional reconstruction loss (\(\mathcal{L}_{\text{rec}}\)) and a regularization loss (\(\mathcal{L}_{\text{reg}}\)). \(\mathcal{L}_{\text{reg}}\) penalizes \(\boldsymbol{y}\) when it closely matches \(\boldsymbol{y}^*\) for impaired speech frames identified by the phone classifier, hence alleviates the impact of articulation disorder presented in the training samples.
\vspace{-0.6em}

\subsection{Consistency loss}
The articulation impairment model described in the preceding section assumes unimpaired articulation in the generated speech \(\boldsymbol{y}_{1:T}\). However, such assumption is not enforced explicitly in the training process. To address the issue, we introduce another loss term inspired by the observation-prediction consistency constraint as described in \cite{reconstruction}. In our research, ``observations'' refers to the input phonemes, and ``prediction'' is the generated speech. We leverage the phone classifier to quantify the inconsistency between input phonemes and generated speech.

Let \(\boldsymbol{x}'_{1:T}\) denote the time-expanded phoneme sequence, aligned with the generated speech \(\boldsymbol{y}_{1:T}\). From the classifier's perspective, which only has access to the generated speech (i.e., the ``prediction''), \(P_{\text{cls}}(\boldsymbol{x}'_{1:T} | \boldsymbol{y}_{1:T})=\prod_{t=1}^T p_t\), where \(p_t\) represents the posterior of the \(t^{\text{th}}\) phoneme label estimated by the classifier. However, according to the TTS model, \(P_{\text{tts}}(\boldsymbol{x}'_{1:T} | \boldsymbol{y}_{1:T})=1.0\) as \(\boldsymbol{x}'_{1:T}\) is what it observes (i.e., the ``observation"). The consistency loss is defined as the KL divergence between the two probability distributions:
\begin{align}
\mathcal{L}_{\text{consis}} = D_{KL}\left(P_{\text{tts}}\Vert P_{\text{cls}}\right) = -\sum_{t=1}^T\log p_t
\label{equation:consis}
\end{align}
Minimizing the consistency loss enhances the consistency between input phonemes and the generated speech, thereby promoting correct articulation in the synthesized output.

The total loss for training the TTS acoustic model is then defined as equation (\ref{equation:augrec}), where \(\beta>0\) and \(\gamma>0\) are hyper-parameters to be tuned.
\begin{align}
\mathcal{L}_{total} = \mathcal{L}_{\text{rec}} + \beta\cdot\mathcal{L}_{\text{reg}} + \gamma\cdot\mathcal{L}_{\text{consis}}
\label{equation:augrec}
\end{align}
\end{spacing}

\vspace{-1.0em}

\section{Experiments}
\begin{spacing}{0.98}
We validate the proposed approach on a real case: creating a synthetic voice for a female Cantonese speaker from recordings of articulation-impaired speech. The target speaker underwent partial-glossectomy six years ago and around 3/4 of her tongue was removed surgically. Phonemes including the velar consonants \wt{k}, \wt{g}, \wt{N}, alveolar consonants \wt{t}, \wt{d} and high vowels \wt{9}, \wt{y} are noticeably impacted in her speech. The quality of the synthetic voice is evaluated in terms of intelligibility, overall impression and speaker similarity. Experimental details are described below.
\vspace{-0.5em}

\subsection{Systems, speech corpora and training details}
We adopt the Mel-spectrogram prediction network from Tactotron 2\cite{tacotron2} for acoustic modelling, but replace the attention module with the duration predictor and aligner from DurIAN \cite{durian} to improve model robustness. We use a pre-trained HiFi-GAN \cite{hifigan} as the vocoder. The external phone classifier is adapted from Jasper\cite{li2019jasper}, with the convolution stride removed to enable frame-wise prediction. We refer to the proposed system, which trains the acoustic model with augmented reconstruction loss, as \textbf{DuriTaco+AugRecLoss}.

The proposed system is compared with two baselines. Both share the same network architecture as the proposed system, but differ in terms of the training objective and/or training data. The first baseline, \textbf{DuriTaco}, trains the acoustic model with conventional \(L_2\) Mel-spectrogram reconstruction loss. The second baseline, \textbf{DuriTaco+VC}, trains the acoustic model on VC generated speech, also with \(L_2\) reconstruction loss. We choose NANSY \cite{nansy} as the VC model for its strong performance in cross-lingual setting. Our assumption is that impaired speech can be viewed as a distinct language, therefore a strong cross-lingual VC model should also be effective for voice conversion between normal and impaired speech. 

Our target speaker has contributed a total of 377 recordings, which gives approximately 24 minutes of speech data. Speech was recorded using a TASCAM DR-44WL at 44.1 kHz in a reasonably quiet environment. We refer to this collection as \textbf{Recording\_T}. We additionally collected 377 recordings of unimpaired speech from another female Cantonese speaker. This speech corpus, named \textbf{Recording\_S}, contains exactly the same content as Recording\_T, and serves as the source speaker's data for voice conversion. For TTS pre-training, we use the multi-speaker Cantonese speech corpus \textbf{CUSENT}\cite{lee2002spoken}. It consists of around 20-hour read speech from 80 speakers, sampled at 16 kHz. The external phone classifier and the VC model are also trained using this corpus.

Speech data are resampled to 22.05 kHz to be consistent with HiFi-GAN's settings for Mel-Spectrogram computation. Acoustic models of the two baselines are pre-trained on CUSENT for 600 epochs at batch size 32, then fine-tuned on the target speaker's data (either original recordings or VC generated speech) for 5000 steps at batch size 16. The acoustic model in the proposed system is initialized with the DuriTaco baseline. It is then fine-tuned on the target speaker's data, combined with 2000 randomly selected recordings from CUSENT, for 750 steps at batch size 32. We empirically find that incorporating multi-speaker speech data helps to protect the phone classifier from adversarial attacks by the TTS model. Hyper-parameters in the loss function are set to \(\beta=0.05, \gamma=0.3, \lambda=25.0\). For all the systems, the fine-tuning process utilized only 347 recordings from the target speaker, with the remaining 30 sentences held out for evaluation. During fine-tuning, parameters of the encoder are frozen. 
\vspace{-0.2em}

\subsection{Objective evaluation}
For objective evaluation, we utilized a separately trained CTC-based automatic speech recognition (ASR) model. It is implemented following the recipe from SpeechBrain\footnote{\url{https://github.com/speechbrain/speechbrain/tree/develop/recipes/TIMIT/ASR/CTC}}, and is trained on a commercial Cantonese speech corpus, which contains 80 hours of read speech from 136 speakers. 

We resynthesized all 377 sentences in Recording\_T with the proposed system and the two baselines. The Phone Error Rate (PER\%) results in Table \ref{tab:intelligibility} show that the DuriTaco baseline largely retains the problematic articulation style observed in the original recordings. On the other hand, the proposed system is able to generate highly intelligible speech, with articulation quality comparable to the VC baseline and unimpaired natural speech. Results in the last two rows suggest that applying both \(\mathcal{L}_{\text{reg}}\) and \(\mathcal{L}_{\text{consis}}\) is the most effective, while applying \(\mathcal{L}_{\text{consis}}\) accounts for most of the improvement. This is expected because \(\mathcal{L}_{\text{consis}}\) explicitly enforces the consistency between the input phonemes and the generated speech.

\begin{table}
		\caption{PER(\%) results on both the original recordings and the resynthesized speech.}
		\label{tab:intelligibility}

		\centering
		\renewcommand{\arraystretch}{0.95}
            \small
 		\begin{tabular}{lc}
			\toprule
			Case  & PER\%$\downarrow$\\
			\midrule
                \midrule
			Recording\_S & 11.2\\
                Recording\_T & 43.1\\
                \midrule
                DuriTaco &   37.6\\
                DuriTaco + VC &  \textbf{14.9} \\
                DuriTaco + AugRecLoss (w/o \(\mathcal{L}_{\text{reg}}\)) & \textbf{13.8} \\
                DuriTaco + AugRecLoss (w/  \(\mathcal{L}_{\text{reg}}\))  & \textbf{12.0} \\
			\bottomrule
		\end{tabular}
  \vspace{-0.5em}
	\end{table}

 \vspace{-0.2em}
 \subsection{Subjective evaluation}
The proposed TTS system and the two baselines were used to generate speech for the 30 sentences in the held-out evaluation set. For each sentence, the three synthetic voices along with the two real speech samples (one from the source speaker and one from the target speaker) formed five stimuli in total. 

We then carried out a web-based listening test to assess the quality of different systems using these stimuli. The test included two parts: a mean opinion score (MOS) test and a similarity mean opinion score (SMOS) test, each consisting of 30 questions corresponding to the test sentences. In each question, a stimulus was randomly chosen from the five systems, and for the SMOS test, a reference recording sample from the target speaker was also provided. Listeners were asked to rate the overall impression or speaker similarity on a 5-point scale. A total of 66 native Cantonese speakers participated in the test, and after filtering out responses with overall impression ratings below 4.0 for natural speech recordings (Recording\_S), we obtained a total of 57 valid responses.

The MOS results in Table \ref{tab:similarity} show that all systems, including the DuriTaco baseline, produce speech that is more natural than the original articulation impaired speech. This general improvement might be attributed to the multi-speaker pre-training, which incorporates the natural articulation style of unimpaired speech into the model. Nevertheless, the VC baseline and the proposed system exhibit the most noticeable improvement in terms of overall impression. On the other hand, the SMOS results demonstrate the superiority of the proposed system over the VC baseline in preserving the target speaker's individuality, which is a desirable quality in the context of personalized speech synthesis. Readers are encouraged to visit the demo page\footnote{\url{https://myspeechproject.github.io/ArticulationRepair/}} to listen to audio samples. 

\begin{table}
		\caption{MOS and SMOS evaluations with 95\% confidence intervals for different systems.}
		\label{tab:similarity}

		\centering
		\renewcommand{\arraystretch}{0.95}
            \small
 		\begin{tabular}{lcc}
			\toprule
			Case & MOS $\uparrow$& SMOS $\uparrow$\\
			\midrule
                \midrule
			Recording\_S & $4.76\pm 0.08$ & $1.06\pm 0.04$\\
                Recording\_T & $2.38\pm 0.12$ & $4.23\pm 0.07$\\
                \midrule
                DuriTaco & $2.88\pm 0.12$ & $\mathbf{4.22\pm 0.15}$\\
                DuriTaco + VC & $\mathbf{3.60\pm 0.09}$ & $3.30\pm 0.19$ \\
                DuriTaco + AugRecLoss & $\mathbf{3.66\pm 0.08}$ & $\mathbf{4.00\pm 0.17}$ \\
			\bottomrule
		\end{tabular}
  \vspace{-0.5em}
	\end{table}
 \end{spacing}

 \vspace{-0.2em}
 
\section{Conclusion}
\begin{spacing}{0.98}
We introduced an augmented reconstruction loss for developing personalized synthetic voices from articulation-impaired speech. 
Experimental results from a real patient case demonstrate that when trained with the proposed loss function, the synthetic voice is able to restore normal articulation in the synthesized output, while effectively maintaining the target speaker's voice identity and speaking style. Although initially validated on post-glossectomy speech, the proposed approach should be applicable to other types impaired speech where the speech disorder mainly affects articulation.
\end{spacing}

\vspace{-0.2em}
\section{Acknowledgements}
\begin{spacing}{0.97}
\vspace{-0.2em}
We would like to extend our sincere gratitude to the target speaker, who generously agreed to showcase her synthetic voice on the demo page. 
\end{spacing}

\bibliographystyle{ieeetr}
{\bibliography{refs}}

\end{document}